## RESEARCH ARTICLE

# Intelligent Backhaul Link Selection for Traffic Offloading in B5G Networks

ANTÓNIO J. MORGADO[1], FIROOZ B. SAGHEZCHI[2], (Senior Member, IEEE),
PABLO FONDO-FERREIRO[3], FELIPE GIL-CASTIÑEIRA[3],
AND JONATHAN RODRIGUEZ[1,4], (Senior Member, IEEE)
[1]Faculty of Computing, Engineering and Science, University of South Wales, CF37 1DL Pontypridd, U.K.
[2]Chair for Distributed Signal Processing, RWTH Aachen University, 52062 Aachen, Germany
[3]Information Technologies Group, atlanTTic Research Center, University of Vigo, 36310 Vigo, Spain
[4]Instituto de Telecomunicações, 3810-193 Aveiro, Portugal

Corresponding author: António J. Morgado (antonio.dasilvamorgado@southwales.ac.uk)

This work was supported in part by the Fundação para a Ciência e Tecnologia (FCT), I.P., under Project UIDB/50008/2020; and in part by Xunta de Galicia (Spain) under Grant ED481B-2022-019.

**ABSTRACT** Fifth Generation (5G) mobile networks considers an expansive set of heterogeneous services with stringent Quality of Service (QoS) requirements, and traffic demand with inherent spatial-temporal distribution, which places the backhaul network deployment under potential strain. In this paper, we propose to harness network slicing, Integrated Access and Backhaul (IAB) technology coupled with satellite connectivity to build a dynamic wireless backhaul network that can provide additional backhaul capacity to the base stations on demand when the wired backhaul link is temporarily out of capacity. To construct the network design, Deep Reinforcement Learning (DRL) models are used to select, for each network slice of the congested base station, an appropriate backhaul link from the pool of available IAB and satellite links that meets the QoS requirements (i.e., throughput and latency) of the slice. Simulation results show that around 20 episodes are sufficient to train a Double Deep Q-Network (DDQN) agent, with one fully-connected hidden layer and Rectified Linear Unit (ReLU) activation function, that adjusts the topology of the backhaul network.

## I. INTRODUCTION

The increasing 5G traffic demand leads to significant variation in the network spatial-temporal domains [1]. Hence, from the spectrum management perspective, it is very instrumental to equip 5G networks with machine learning tools to identify spectrum availability, in particular to perceive the degree of occupancy within the deployed network for providing alternative wireless backhaul links through exploiting integrated access and backhaul (IAB) base stations and available satellite links for offloading traffic. Figure 1 illustrates the scenario under study. Here, the goal is to activate, deactivate, or configure the terrestrial (orange) and satellite (blue) backhaul links so that they can deliver the required capacity to specific base stations, while they do not interfere with other services using the same bands (links that are depicted with shadow).

IAB is considered as a means to reduce deployment costs in 5G networks and beyond, especially in ultra-dense scenarios such as mmWave networks [2]. The main challenges in IAB relate to the self-configuration of the network and a traffic path selection for every network slice in order to optimize the network performance and guarantee the desired Quality of Service (QoS).

Most of the existing works on IAB focus on radio resource allocation, or backhaul path selection as can be seen in Table 1. Deep Reinforcement Learning (DRL) techniques have been explored for addressing these challenges. For instance, a DRL-based radio resource management solution for congestion avoidance was proposed in [3]. The authors

The associate editor coordinating the review of this manuscript and approving it for publication was Bilal Khawaja.

**TABLE 1.** Related works.

| Reference | System model | | | | Radio resource management goals | | | | | |
|---|---|---|---|---|---|---|---|---|---|---|
| | DRL | IAB | NTN (Satellite/UAV) | Network slicing | Backhaul path selection | Spectrum allocation | Power control | Delay | Coverage | Capacity |
| [2] | | √ | | | √ | | | | | √ |
| [3] | √ | √ | | | | √ | | | | √ |
| [4] | √ | √ | | | | √ | | | | √ |
| [5] | √ | √ | | | | √ | √ | | | √ |
| [6] | √ | √ | | | √ | | | √ | | |
| [7] | √ | √ | | | √ | | | | | √ |
| [8] | | √ | √ (Satellite/UAV) | | | | | | | |
| [9] a | √ | √ | √ (Satellite/UAV) | | √ | √ | √ | √ | | √ |
| [10] | | √ | √ (UAV) | | | | | | √ | |
| [11] | | √ | √ (UAV) | | | | | | | √ |
| [12] | | √ | √ (UAV) | | | | | | | √ |
| [13] | | √ | √ (UAV) | | √ | | | | | √ |
| [14] | | √ | √ (Satellite) | | √ | | | | | √ |
| [15] | | √ | √ (Satellite) | | | √ | | | | √ |
| [16] b | | √ | √ (Satellite) | | | √ | | | | |
| [17] | | √ | √ (Satellite) | | | | √ | | | √ |
| [18] | | √ | √ (Satellite) | | √ | | | √ | | √ |
| [19] | | √ | √ (Satellite/UAV) | | √ | | | | | √ |
| [20] | √ | √ | √ (UAV) | | | √ | √ | | | √ |
| Our work | √ | √ | √ (Satellite) | √ | √ | | | √ | | √ |

a Reference [9] is a survey paper, so the several topics addressed in the table originated from different works, some of them (e.g., on NTN topic) do not use machine learning algorithms.
b Reference [16] is a game theory paper that intends to design a pricing mechanism to motivate the terrestrial and the satellite operators to offload terrestrial traffic to the satellite backhaul network.

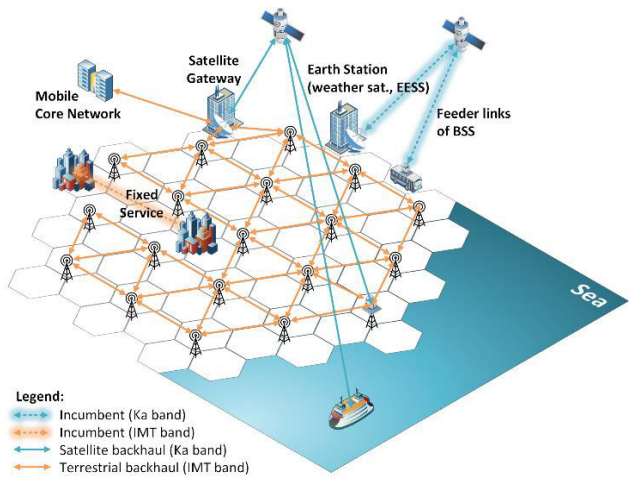

**FIGURE 1.** Dynamic wireless backhaul network.

of [4] used DRL techniques for spectrum allocation with the aim of maximizing the sum log-rate of the users. DRL was also used in [5] for jointly addressing the spectrum allocation and power control in IAB networks. The authors of [6] proposed a DRL-based cross-layer approach for jointly tackling routing and radio resource allocation in multi-hop IAB scenarios. In [7], the authors propose a multi-agent DRL framework to optimize user throughput in an IAB network. The resource optimization deals with routing paths and scheduling of directional transmissions along established links, coordinating both access and backhaul transmissions to maximize the downlink throughput observed by the UEs.

The integration of Non-Terrestrial Networks (NTN) with mobile networks is considered since 3GPP Release 17 [8].

A potential application of NTN technologies in mobile networks is its integration in the IAB architecture to increase the network capacity [9]. The works on this topic are focused on two main technologies: Unmanned Aerial Vehicles (UAV) and satellite communications. The main challenge of the studies addressing UAV-assisted IAB networks relate to the trajectory planning of UAVs acting as flying base stations in the three-dimensional space in order to optimize different Key Performance Indicators (KPIs). For example, the authors of [10] used ray tracing techniques to improve coverage. In [11], the authors jointly reduced the number of UAVs while increasing transmission rate, and in [12], the authors focused on maximizing the sum rate. Other works, such as [13], focused on the path selection strategies for UAV-assisted multi-hop wireless communications scenarios.

Satellite-terrestrial IAB networks allow increased coverage ranges at lower costs. The intrinsic high latencies of satellite links make them a potential candidate for traffic offloading in delay tolerant services [14]. In [15], the authors addressed spectrum allocation in such networks with the aim of maximizing the sum rate, assuming that both satellite and terrestrial networks can reuse the same frequency bands. Other approaches, such as [16], used game theory to propose a mechanism for traffic offloading to Low Earth Orbit (LEO) satellites, based on a Stackelberg game. The authors in [17] elaborate on the standardization compatibility of using NTNs in IAB architectures, showing its feasibility and analyzing a case study using LEO satellites. The authors in [18] propose a dynamic backhaul network reconfiguration, including smart antennas, dynamic routing and load balancing in satellite-assisted mobile networks. Their proposal reduces the congestion by offloading traffic to satellite links when needed. The authors in [19] consider also the exploitation of UAVs in satellite-assisted mobile networks and formulated the problem of user association and resource allocation as a competitive game problem. Finally, the authors of [20] propose a distributed DRL approach for joint resource allocation and optimal deployment of UAVs to maximize the total sum rate of an IAB network.

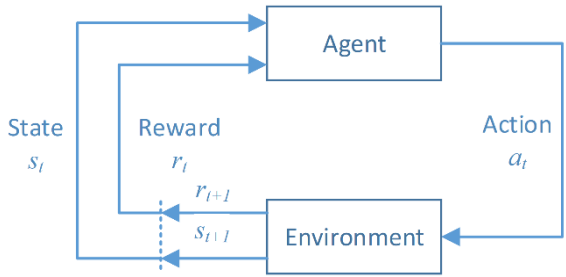

**FIGURE 2.** Interaction between the agent and the environment [21].

Despite these contributions, most of the related works in the literature focus exclusively on the allocation of radio resources to a single network slice. However, in this paper, we shift our focus to the selection of an appropriate backhaul link – among a pool of existing wired, wireless (IAB), or satellite links – for every network slice to make sure that the QoS requirements of all slices are met. In this context, our main contributions are summarized as follows:

- We design a DRL agent that selects an appropriate backhaul link for every network slice, in contrast to existing works, which mostly address radio resource allocation to a single network slice.
- For every slice, we select a backhaul link from a pool of wired, wireless (IAB) and satellite links.
- For every allocated backhaul link, our DRL agent verifies the compliance of the selected link with the QoS requirements (i.e., throughput and latency) of the served slice.

The remainder of this paper is organized as follows. Section II reviews the required background on DRL and IAB. Section III describes our system model along with our proposed DRL-based backhaul link selection algorithm. Section IV presents our simulation results. Finally, Section V concludes this paper and draws some guidelines for future work.

## II. ENABLING TECHNOLOGIES
### A. DEEP REINFORCEMENT LEARNING
Reinforcement Learning (RL) is a branch of machine learning that learns from explorative interactions and reward collections. It consists of an environment and an agent that interacts with that environment at discrete time steps, executing over it a sequence of actions to maximize the discounted cumulative reward.

At time instant $t$, the agent observes the state $s_t$, which is a collection of parameters that characterize the current status of the environment. Based on the state $s_t$, the agent selects the action $a_t$ to be executed by the environment, causing the transition of the environment from state $s_t$ to state $s_{t+1}$. After executing the action $a_t$, the environment returns the agent an instant reward $r_t$, which is a feedback that the agent receives evaluating numerically the quality of the action, using which the agent can distinguish a good action from a bad one.

The RL agent selects actions based on a policy $\pi$, that maps states to actions, i.e., $\pi(s, a)$ represents the probability of the agent selecting action $a$ when the environment is in state $s$.

As mentioned before, the goal of the agent is to maximize the *discounted cumulative reward* [21]:

$$R_t = r_{t+1} + \gamma r_{t+2} + \gamma^2 r_{t+3} + \ldots = \sum_{k=0}^{k=\infty} \gamma^k r_{t+k+1} \quad (1)$$

where $r_t$ is the *instantaneous reward* and $\gamma$ ($0 \leq \gamma \leq 1$) is the *discount rate*. For that, the agent may have to estimate the *state-action value function*, $Q^\pi(s, a)$, also known as *critic function*, that expresses the expected return when the environment is in state $s$ and the agent executes action $a$ and afterwards follows policy $\pi$ [21].

$$Q^\pi(s, a) = \mathbb{E}_\pi \{R_t \mid s_t = s, a_t = a\}$$
$$= \mathbb{E}_\pi \left\{ \sum_{k=0}^{k=\infty} \gamma^k r_{t+k+1} \mid s_t = s, a_t = a \right\} \quad (2)$$

The optimal value function, i.e. the $Q^\pi(s, a)$ that maximizes (1), is defined as [21]

$$Q^*(s, a) = \max_\pi Q^\pi(s, a) \quad (3)$$

which may also be expressed by the Bellman equation [21]

$$Q^*(s, a) = \mathbb{E} \left\{ r_{t+1} + \gamma \max_{a'} Q^*(s_{t+1}, a') \mid s_t = s, a_t = a \right\} \quad (4)$$

$$= \sum_{s'} P_{ss'}^a \left[ R_{ss'}^a + \gamma \max_{a'} Q^*(s', a') \right] \quad (5)$$

where $P_{ss'}^a$ depicts the probability of transition to state $s'$ when the environment is in state $s$ and the agent executes action $a$ on it and $R_{ss'}^a$ is the corresponding reward that the agent receives. The task of the RL agent is to solve the Bellman equation and determine the $Q^*(s, a)$ that maximizes the *discounted cumulative reward* defined in (1). However, most of the RL methods do not compute $Q^*(s, a)$ using (5) directly. Instead, they try to iteratively approximate the solution of the Bellman equation (4) without requiring perfect knowledge of the transition probabilities $P_{ss'}^a$ of the environment (model free RL) [23]. In addition, when the observation and/or the action space have too many elements, it is infeasible to represent the Q-function using a lookup table, so it has to be represented by a function approximator, e.g. a neural network [21], [22].

The Q-learning algorithm was one of the first model free RL algorithms proposed in the literature [23], [24]. The idea behind Q-learning is to iteratively approximate the critic function $Q(s,a)$ to the optimum value given by Bellman equation (4).

$$Q(s,a) \leftarrow Q(s,a) + \alpha \left[ r_{t+1} + \gamma \max_{a_{t+1}} Q(s_{t+1}, a_{t+1}) - Q(s,a) \right]$$

$$Q(s,a) \leftarrow (1-\alpha) Q(s,a) + \alpha \left[ r_{t+1} + \gamma \max_{a_{t+1}} Q(s_{t+1}, a_{t+1}) \right] \quad (6)$$

Note that in (6), we have dropped the expected value operation E{.} for simplicity.

Theoretically, Q-learning converges to the optimum $Q(s,a)$ as the algorithm iterations approach infinity [24]. However, in (4), when we use a neural network to model the critic function $Q(s,a)$, we evaluate the expected value of the critic function $Q(s,a)$ using weights from a previous iteration [26], which may lead to oscillations or divergence of the $Q(s,a)$.

A variant of Q-learning, specifically designed to approximate the $Q(s,a)$ function using neural networks was introduced in [25] and improved in [26]. It is called Deep Q-Network (DQN) and introduces the following novelties [25], [26] when compared to Q-learning:

- It uses an *experience replay buffer* to store the experience in each time step $(s_t, a_t, r_t, s_{t+1})$. From this buffer random samples are selected in each time step, thus providing uncorrelated data to train the neural network.
- It uses *mini-batches* of random samples selected from the experience replay for efficient computation of the gradients to update the weights of the neurons.
- It employs a second neural network, called the 'target critic', so the target reward values given by Bellman equation can be updated less frequently, with a periodicity of a given number of time steps.

These novelties alleviate the problem of correlated data and non-stationary distributions, smooth the learning and avoid oscillations or divergence in neural network weights [25], [26].

In [27], the authors proved that using the same critic in DQN to calculate the target reward $y_i$ and to select the action $a'$ leads to an overoptimistic $y_i$. To reduce this bias, they propose to use two neural networks, one for online critic to infer the best action and the other one for target critic to evaluate the selected action. Algorithm 1 shows the DDQN model, which is essentially a DQN model, where its step 3.2.5 has been divided into two sub-steps.

Double DQN, is shown to reduce the overoptimistic $y_i$, thus resulting in more stable and reliable learning, which allows to find better policies [27].

### B. INTEGRATED ACCESS AND BACKHAUL (IAB)
IAB is a 5G feature that enables the base stations with wired backhaul, called IAB-donors, to provide wireless backhaul links to other base stations that lack wired backhaul links, called IAB-nodes (see Fig. 3).

**Algorithm 1** Double DQN (DDQN)

Define parameters:
$\alpha$ (learning rate), $\varepsilon$ (initial value of the probability to select a random action), $\varepsilon_{decay}$ ($\varepsilon$ decay in each timestep), $\gamma$ (discount factor), $\mathcal{A}$ (discrete action space), N (size of experience replay buffer), M (size of the mini-batch), C (number of timesteps between updates of the target critic)

1. Create an experience replay buffer $D$ with capacity for storing $N$ experiences $(s, a, r, s')$.
2. Initialize the on-line critic $Q(s, a, \phi)$ with random weights $\phi$, and initialize the target critic $Q_{target}(s, a, \phi_{target})$ with the same weights, i.e., $\phi_{target} = \phi$.
3. For each training episode:
   3.1 Get initial observation $s$ from the environment.
   3.2 For each timestep of the episode:
   3.2.1 Select a random action $a$ with probability $\varepsilon$; otherwise, select the action $a$ that maximizes the current value of $Q(s, a, \phi)$, i.e., $a = \max_{a \in \mathcal{A}} Q(s, a, \phi)$.
   3.2.2 Execute action $a$. Observe the instant reward $r$ and the next state $s'$.
   3.2.3 Store the acquired experience $(s, a, r, s')$ in the experience replay buffer $D$.
   3.2.4 Sample a random minibatch of $M$ experiences $(s_i, a_i, r_i, s'_i)$ from the experience replay buffer $D$.
   3.2.5 For each sample of the minibatch: if $s'_i$ is a terminal state, set the Q-function target value $y_i$ to $r_i$. Otherwise, apply Double Q-learning as follows:
   3.2.5.1 Use the on-line critic $Q(s, a, \phi)$ to select the action $a'$:
   $$a' = \max_{a \in \mathcal{A}} Q(s'_i, a, \phi)$$
   3.2.5.2 Use the target critic $Q_{target}(s, a, \phi_{target})$ to compute the target reward $y_i$, i.e.
   $$y_i = r_i + \gamma \cdot Q_{target}(s'_i, a', \phi_{target})$$
   3.2.6 Compute the loss function across all samples of the minibatch:
   $$L = \frac{1}{M} \sum_{i=1}^{M} (y_i - Q(s_i, a_i, \phi))^2$$
   3.2.7 Compute the gradients $\Delta \phi$ of the loss function with respect to the weights of the neural network:
   $$\Delta \phi = \frac{1}{2} \nabla_\phi (L)$$
   3.2.8 Use stochastic gradient descent to update the neural network weights based on the computed gradients, so the neural network output approximates $y$, i.e.,
   $$\phi = \phi + \alpha \cdot \Delta \phi$$
   3.2.9 Every $C$ timesteps, update the weights of the target critic, i.e., $Q_{target} = Q$.
   3.2.10 Set the observation $s$ to $s'$.
   3.2.11 Set $\varepsilon = \varepsilon \cdot (1 - \varepsilon_{decay})$.

The development of IAB was motivated by the need for deploying dense mmWave 5G networks at a lower cost than using fiber. IAB is economically viable in 5G due to the wider bandwidth of mmWave, the support of beamforming and massive MIMO, which allow to build a wireless backhaul network with performance similar to fiber. The main use cases envisioned for IAB are coverage extension, deployment of outdoor small cells, and fixed wireless access (FWA) to indoor hotspots.

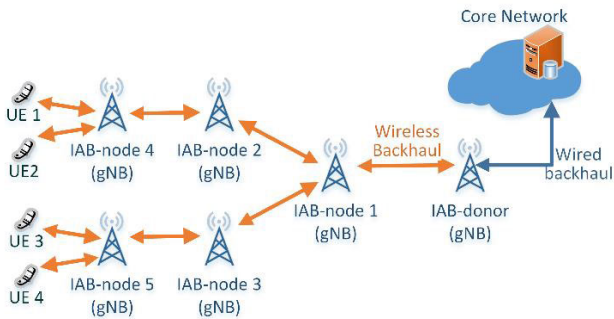

**FIGURE 3.** A possible IAB topology.

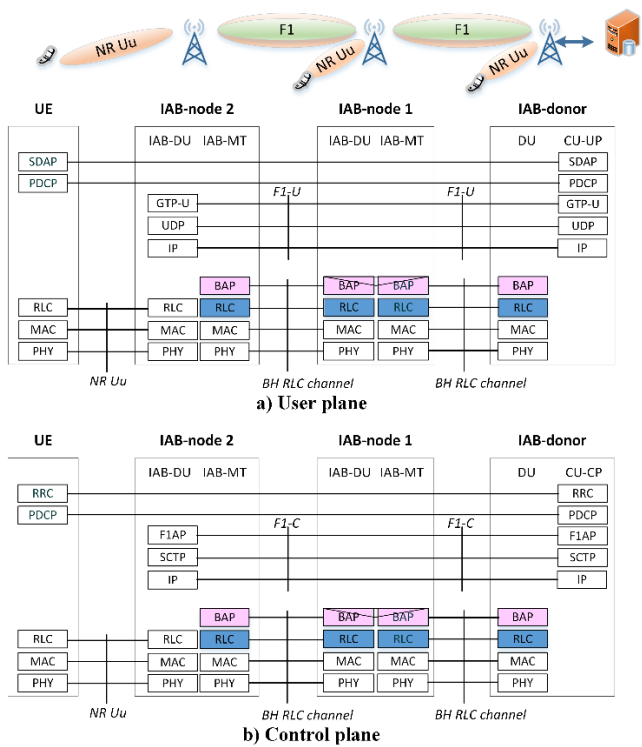

**FIGURE 4.** IAB protocol stack (based on [28], [29]).

Normalization of IAB started as a study item in 3GPP Release 15 [28], although IAB is a feature of the 3GPP 5G standard first introduced in Release 16 [29], and improved in Release 17 and 18. The main features of IAB in Release 16 are:

- Support of multi-hop backhauling for flexible coverage extension;
- Support for QoS differentiation and enforcement for allowing the transmission of all the 5G QoS classes;
- Support of network topology adaptation and redundant connectivity (i.e. the same data is transmitted over different paths) for optimal and robust operation of wireless backhaul in the presence of mmWave signal blockages;
- Support of in-band and out-of-band backhauling to enable access and backhaul wireless links to share the same carrier frequency or use different frequencies, respectively. In case of in-band backhauling, the IAB nodes cannot receive and transmit at the same time unless high isolation exists between the receiver and the transmitter of the IAB nodes;
- Dynamic scheduling for fast adaptation of the resources allocated to the access and backhaul networks, in case of in-band backhauling [30];
- Transparency to the UE, i.e., no additional features need to be deployed in the UEs, so legacy terminals can access the network seamlessly.

Release 17 added the following IAB enhancements [31]:

- Inter-donor migration, that allows an IAB node to migrate from one IAB donor to another;
- Inter-donor topological redundancy, that allows an IAB node to connect to two different IAB-donors using dual connectivity.

Release 18 added the following further enhancements [32]:

- Mobile IAB nodes: these are IAB-nodes mounted in vehicles to provide 5G coverage to onboard and/or surrounding UEs. These mobile IAB nodes do not connect to other IAB nodes, only UEs. However, they can migrate within the same IAB-donor or to a different IAB-donor.

These features are achieved by a protocol stack as depicted in Fig. 4. The IAB-donor is divided in CU and DU parts as traditional gNBs, while the IAB node is divided in a mobile termination (MT) part, which is used to communicate with a parent node, and a DU part, used to communicate with child IAB nodes or with normal UEs. Each IAB node has an IP address, which is routable from the CU of the IAB-donor. For efficient multi-hop packet forwarding between several IAB nodes, a backhaul adaptation protocol (BAP) is also introduced in every DU and MT modules. The objective of this protocol stack is to create hop-by-hop RLC channel between the IAB nodes in order to achieve faster single-hop retransmission. However, for in-band backhauling, unless high isolation exists between the MT and DU of the IAB nodes, the IAB node cannot transmit and receive at the same time.

Besides routing decisions, since the CU has an overview of the whole backhaul path, it can also be used for other centralized procedures such as handover decisions, modification of the topology of the backhaul network, bearer mapping, etc. In our case, we are interested in changing the backhaul links from time to time. For this, the CU of the IAB-donor will have to change the UL and DL routing tables of the IAB node(s) each time it needs to adapt the topology of the UL or DL backhaul networks.

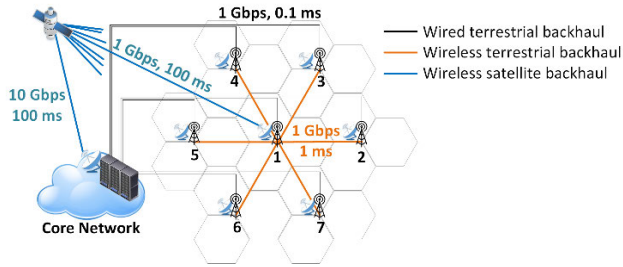

**FIGURE 5.** IAB scenario under study.

## III. SYSTEM MODEL
### A. SCENARIO
We construct and evaluate RL models for the scenario shown in Fig. 5. In this scenario, all seven 5G base stations (gNB) have a wired and a satellite backhaul connecting them to the core network, and they may also use IAB to borrow additional backhaul capacity from the neighbor base stations.

We assume that the congested base station, i.e., the base station needing to use satellite backhaul or to borrow wireless backhaul from the neighbor base stations is *base station no.1* (BS1).

In this scenario, the objective is to configure the network topology to meet the QoS requirements of the $N_s$ slices to be served by BS1, considering the traffic load in the remaining base stations. We describe each of these configuration parameters and constraints in the next numbered paragraphs.

#### 1) NETWORK TOPOLOGY
We configure the backhaul network to have $N_{BS}+2$ nodes, where $N_{BS}$ nodes (nodes $1\ldots N_{BS}$) are the 5G base stations, and the remaining nodes (node 0 and node 99) are the core network and the satellite, respectively. Each node is characterized by a name and its geographical location. The nodes are connected together by directional links represented by the tuples:

$$\text{Link}(x, i, j, b, d)$$

where $x \in \{\text{wired, wireless, satellite}\}$ represents the link type, $i \in \{0, 1, \ldots, N_{BS}, 99\}$ represents the source node and $j \in \{0, 1, \ldots, N_{BS}, 99\}$ represents the target node, $b$ represents the total bandwidth of the link, and $d$ represents the latency of that link.

In the case of the wired links, we assume there are 2 fibers, each one to transmit traffic in one direction (DL, UL). In the case of wireless and satellite links, we assume that none of them interfere with any incumbent network that might already be operating in the same bands.

The characteristics of the links are presented in Table 2.

We stress the fact that the wireless connections may have a bandwidth of 1 Gbps (DL) / 1 Gbps (UL) for backhauling when their entire bandwidth is available to backhaul links, i.e. they have no UE in their cells. On the contrary, when the cells are heavily loaded, part of these bandwidths are occupied by the access network traffic and the remaining part by the backhaul traffic. Moreover, the baseline delays indicated in Table 2 may increase by the queuing delay [33].

**TABLE 2.** Characteristics of the links.

| | Bandwidth | | Delay | |
|---|---|---|---|---|
| | DL | UL | DL | UL |
| Wired backhaul | 1 Gbps | 1 Gbps | 0.1 ms | 0.1 ms |
| Satellite backhaul | 1 Gbps | 1 Gbps | 100 ms | 100 ms |
| Wireless backhaul | Up to 1 Gbps | Up to 1 Gbps | 1 ms | 1 ms |

#### 2) SLICE PROFILES IN BS 1
Slice profiles define the variation of the traffic demand, required in each direction (DL, UL), by each of the $N_S$ slices during one day discretized in 15 minute intervals. They also indicate what is the maximum delay required by each slice in each direction during each interval. The $N_S$ traffic profiles are stored in a timetable with the following format:

$$\text{Slice\_profile}(t, i, sid, thdl, thul, ddl, dul)$$

where $t$ is the time, which spans a 24-hour period discretized in 15 minute intervals, $i$ is the BS serving that slice, $sid$ is the slice identifier. For each time interval, $thdl$ and $thul$ represent the throughput required in that time interval by slice $sid$ in DL and UL, respectively, and $ddl$ and $dul$ are the maximum delay that the slice can tolerate in the same time interval.

#### 3) TRAFFIC LOAD OF THE REMAINING $N_{BS}$-1 BASE STATIONS
We defined several traffic load profiles to represent base stations in different situations, i.e., a base station periodically congested, a base station congested in one part of the day, and a non-congested base station. Each profile is stored in a different timetable with the following format:

$$\text{BS\_load\_profile}(t, thdl, thul)$$

where $t$ is the time, defined for a period of 24 hours discretized in 15-minute intervals. For each time interval, the load profile defines the total throughput consumed by the base station in DL and UL respectively to serve its attached mobile users. For the simulation purpose, each of these profiles can then be assigned to any of the $N_{BS}$-1 base stations following any load assignment strategy.

The task of the RL agent is to decide, every 15 minutes, if the BS1 needs to use the satellite backhaul or borrow wireless backhaul capacity from its $N_{BS}$-1 neighbor base stations for any of the $N_S$ slices. To take this decision, the agent will observe the state of the environment, select an action, and receive a reward to give it a notion about how good the selected action was. In the following, we elaborate this further.

### B. ACTION MODEL
The action that the RL agent has to take is to select, every 15 minutes, the $N_S$ backhaul links (UL, DL) for each of the $N_S$ network slices, taking into account the current traffic load of the wired and satellite connections of BS1 and the traffic load of the surrounding base stations (BS2...BS7). The format of the action is shown in Table 3. As we can see from this table, the action space is of discrete type and has size $N_{BS}+1$ since at a given time instant, the agent can select either the wired or satellite backhaul, or one of the six wireless backhaul links offered by BS1's neighbor BSs.

**TABLE 3.** Action format.

| Selectable links for each slice | | | | | | | |
|---|---|---|---|---|---|---|---|
| Satellite | Wired | Wireless | | | | | |
| Satellite Link BS1 | Wired Link BS1 | Wireless Link BS2 | Wireless Link BS3 | Wireless Link BS4 | Wireless Link BS5 | Wireless Link BS6 | Wireless Link BS7 |
| Only one of these links can be TRUE at a time | | | | | | | |

**TABLE 4.** Observation format.

| Slice $s$ requirements | | Free bandwidth (Mbps) | | | | | | | | Delay (ms) | | | | | | | |
|---|---|---|---|---|---|---|---|---|---|---|---|---|---|---|---|---|---|
| Throughput | | Delay | | Sat. | | BS1 | | ... | | BS7 | | Sat. | | BS1 | | ... | BS7 |
| DL | UL | DL | UL | DL | UL | DL | UL | ... | DL | UL | DL | UL | DL | UL | ... | DL | UL |

We assume that UL and DL traffic of each slice is carried by a different fiber of the 'same' backhaul link, i.e., only one backhaul link is selected for every slice. Moreover, the RL agent makes one decision at a time, i.e. to allocate the links for the $N_S$ slices, it performs a sequence of $N_S$ actions.

### C. OBSERVATION MODEL
Our considered observation model is a vector that contains three different parts. In the first part, we include the throughput and delay requirements of the slice being allocated in a given time instant, i.e., the QoS level required by that slice. Following this information, the second part provides the agent with the information about the current bandwidth available in the satellite link of BS1, the wired link of BS1, and in each of the wireless backhaul links of the surrounding base stations (BS2...BS7). Finally, the third part includes the latency that is incurred to connect BS1 with the core network using each of the different available paths, i.e.,

- the satellite link between BS1 and the core network;
- the wired link between BS1 and the core;
- the wireless link connecting BS1 and BS$n$ ($n = 2...7$) followed by the wired link connecting BS$n$ ($n = 2...7$) to the core network.

The format of the observation is illustrated in Table 4. It represents a continuous observation space.

All of Table 4 values are normalized so they all vary in the same interval before feeding them to the neural network.

### D. REWARD MODEL
We adopted the following reward model. The agent receives a reward of +1 when for a given slice, it selects a backhaul link that connects BS1 to the core network meeting the QoS requirements of the slice, so the selected link can be used as the backhaul for the slice under consideration. Otherwise, if the selected link is not capable of being used for the backhaul of the slice under consideration, because of violating the required QoS level, the agent receives a reward 0, and no backhaul link is allocated for this slice.

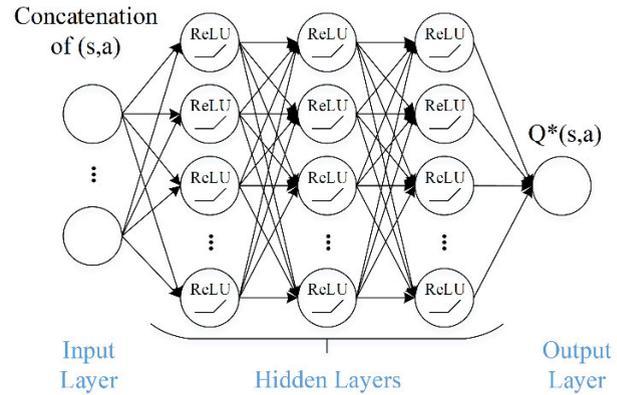

**FIGURE 6.** Fully-connected Q-network used as a critic model.

### E. CRITIC MODEL
We used a Double Deep Q-Network (DDQN) [27] agent to select the backhaul links, since it can work with continuous observation space and discrete action spaces. This model-free, value-based RL agent works by estimating the optimum state-action value function $Q^*(s,a)$, indicated by (4), using a neural network. The optimum policy is then derived by the agent by selecting the action that maximizes $Q^*(s,a)$ for a given state.

The number of elements of the input layer of the neural network must be equal to the sum of the number of elements of the observation (36 parameters, see Table 4) and the number of elements of the action (8 parameters, see Table 3), that is 44 elements in total. We concatenate the observation with the action and provide the concatenated vector as the input to the neural network.

As for the hidden layers, we use fully-connected layers with ReLU (rectified linear unit) activations.

The output is a fully-connected layer with one neuron that estimates $Q^*(s,a)$.

### F. SIMULATOR IMPLEMENTATION
We implemented an RL agent that interacts with a software-defined network (SDN) simulator which acts as the environment. The agent sends commands to the environment whenever it wants to allocate BS1 either a wired backhaul (i.e., a wired link from BS1 to the core network) or a wireless backhaul (i.e., a wireless link from BS1 to BS$n$ with $n = 2...7$, followed by a wired link from BS$n$ to the core network, or a wireless link from BS1 to the satellite followed by a wireless link from the satellite to the core network). Then, the environment reads from the SDN simulator what was the bandwidth and latency effectively allocated for each slice, and computes the reward and the next state to be sent to the RL agent.

**TABLE 5.** DRL hyperparameters.

| Hyperparameter | Value |
|---|---|
| $\alpha$ (learning rate) | 0.001 |
| $\gamma$ (discount factor) | 0.99 |
| $\varepsilon$ (initial exploration probability) | 0.99 |
| $\varepsilon_{decay}$ ($\varepsilon$ decay rate) | 0.01 |
| M (minibatch size) | 64 samples |
| N (size of experience replay buffer) | 20000 samples |
| C (periodic update of the target critic) | 4 timesteps |
| Gradient descent optimization algorithm | adam (adaptive moment estimation) |

**TABLE 6.** Required latency of the network slices.

| | Delay | |
|---|---|---|
| | DL | UL |
| eMBB slice | 100 ms | 100 ms |
| eMTC slice | 10000 ms | 10000 ms |
| uRLLC slice | 1 ms | 1 ms |
| IoT slice | 300 ms | 300 ms |

The SDN backhaul architecture consists of a set of SDN switches co-located along with the base stations, interconnected with different connectivity options. The SDN switches are connected to the SDN controller. An SDN application, running on top of the SDN controller, is responsible for reconfiguring the flow tables of the SDN switches according to the desired forwarding path for each network slice.

The network topology (Fig. 5) is internally represented as a directed graph, with a special node labelled as the core network. Each edge of the graph includes the information of the throughput and latency of the corresponding link.

The throughput measured for a given observation for a network slice will be the requested value by the slice if the slice is allocated. Otherwise, the simulator will return 0 if the slice is not allocated due to lack of sufficient throughput capacity or the violation of the latency requirement of the slice.

On the other hand, the latency is calculated as the sum of the individual latency values of the links in the path up to the core network and also the packet delay caused by the waiting time at each network interface. For the latter, we have modeled all network interfaces as an M/D/1 queue and calculated the average waiting time of each packet at each network interface, following the same approach described in [33].

### G. SIMULATOR VALIDATION

We consider $N_S = 4$ slices to be allocated at BS1, every 15 minutes, taking into account throughput and latency constraints. Thus, throughout the day each slice has to be allocated in 24 hours $\times$ 60 minutes / 15 minutes = 96 occasions. From these $96 \times 4 = 384$ occasions we use $67 \times 4 = 268$ ($\sim$70%) for training, $9 \times 4 = 36$ ($\sim$10%) for cross validation, and $20 \times 4 = 80$ ($\sim$20%) for testing.

Since our reward model assigns a reward +1 when each slice is allocated, and a reward 0 otherwise, the maximum undiscounted episode reward during an episode with 67 timesteps is 4 slices $\times$ 67 timesteps = 268. However, using exhaustive search method, we found that in 6 timesteps, there will be one slice that is impossible to allocate, which means that the highest episode reward that the agent could collect is $268 - 6 = 262$.

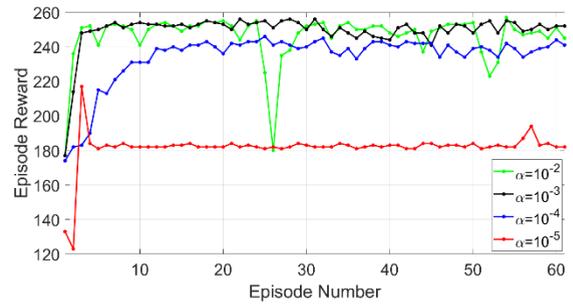

**FIGURE 7.** Tuning the learning rate of the DDQN agent.

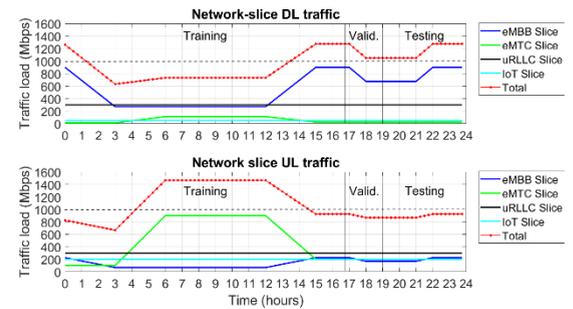

**FIGURE 8.** Daily traffic profiles of the four network slices of BS1.

We trained the agent with different hyperparameters to figure out the best model. For instance, Fig. 7 illustrates the episode reward for different learning rates. At the end, the optimum hyperparameters of the DDQN agent were the ones in Table 5.

## IV. SIMULATIONS
### A. TRAFFIC PROFILES OF THE NETWORK SLICES

For simulations, we used our own constructed synthetic datasets representing the QoS requirements (throughput and delay) of three different types of slices prevalent in mobile networks: 1) an enhanced mobile broadband (eMBB) or enhanced machine type communications (eMTC) slice that needs a high bandwidth during day hours (e.g., in the morning or afternoon) but has a relaxed network latency; and 2) an ultra-reliable low-latency communications (uRLLC) slice that needs constant bandwidth throughout the day but has a stringent network latency (1 ms). Our objective was to examine if the agent was able to learn a good strategy, i.e., selecting lightly loaded BSs for the eMBB or eMTC slices and the fastest link (i.e., the wired link) for the uRLLC slices.

During a typical day, the assumed traffic profiles of the four slices supported by BS1 are illustrated in Fig. 8. As shown, from 04:30 to 14:30, the 1 Gbps wired connection has not enough capacity to support all the UL traffic. The same problem exists in DL from 13:45 to 01:00. In these periods the RL agent has to select the satellite backhaul or one of the neighbor BSs to provide wirelessly the necessary backhaul capacity for some of the slices.

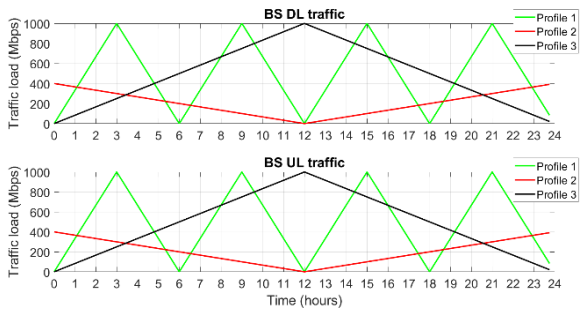

**FIGURE 9.** Daily traffic load of surrounding BSs (BS2-7).

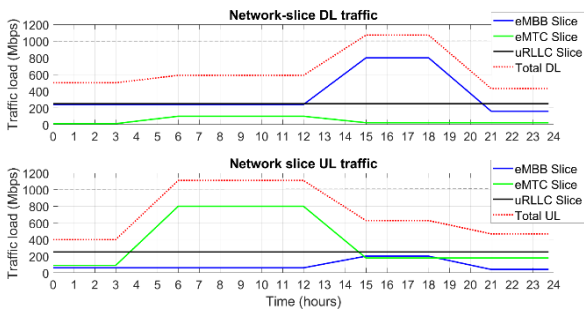

**FIGURE 10.** Daily traffic of the three network slices in BS1 used in simulations without satellite backhaul.

The latency constraints of each slice are shown in Table 6.

### B. TRAFFIC PROFILES OF THE SURROUNDING BS

The surrounding cells (cells BS2 to BS7) also have to support their local traffic (UEs in their cells), in which case they can only lend to BS1 their remaining capacity for backhaul purposes. Therefore, to represent a realistic scenario, we synthetically generated heterogeneous traffic loads for the access network of the surrounding BSs (i.e., highly, lightly, or periodically loaded traffic profiles). The objective was to evaluate if the DRL agent was able to learn to select the BSs that were lightly loaded in every instant or the BSs that were periodically loaded during their low congestion periods. Fig. 9 indicates the load of the surrounding base stations, assuming base station number $b$ was assigned profile $p$ according to:

$$p = ((b-1) \bmod 3) + 1, \quad b = 2, \ldots, 7 \quad (7)$$

As a result, BS4 and BS7 have *profile 1* and are heavy loaded only for small periods of time. BS2 and BS5 have *profile 2* and are lightly loaded, and BS3 and BS6 have *profile 3* and are heavy loaded mainly at noon.

When BS1 needs to borrow wireless backhaul capacity from the neighbor stations, or from the satellite, for any of its four slices, i.e. during the periods 04:30-14:30 and 13:45-01:00, the RL agent has to decide:

- which slice from BS1 is going to have its backhaul traffic transmitted through the satellite or one of the neighbor base stations;

**TABLE 7.** Observation format (No satellite).

| Slice $s$ requirements | | | | Free bandwidth (Mbps) | | | | | Delay (ms) | | | | |
|---|---|---|---|---|---|---|---|---|---|---|---|---|---|
| Throughput | | Delay | | ~~Sat~~ | BS1 | … | BS7 | | ~~Sat~~ | BS1 | .. | BS7 | |
| DL | UL | DL | UL | ~~DL~~ ~~UL~~ | DL UL | … | DL UL | | ~~DL~~ ~~UL~~ | DL UL | .. | DL UL | |

**TABLE 8.** Action format (No satellite).

| Selectable links for each slice | | | | | | | |
|---|---|---|---|---|---|---|---|
| ~~Satellite~~ | Wired | Wireless | | | | | |
| ~~Satellite Link BS1~~ | Wired Link BS1 | Wireless Link BS2 | Wireless Link BS3 | Wireless Link BS4 | Wireless Link BS5 | Wireless Link BS6 | Wireless Link BS7 |
| Only one of these links can be TRUE at a time | | | | | | | |

- which of these nodes is going to transport this backhaul traffic.

### C. RESULTS

#### 1) THE AGENT SELECTS A LINK AMONG TERRESTRIAL LINKS ONLY, BASED ON THROUGHPUT AND LATENCY CONSIDERATIONS

In this situation, as we exclude the satellite link, the observation and action formats were reduced as shown in Table 7 and Table 8, respectively.

In addition, since the satellite was not available for backhaul purposes, we modified the traffic profiles of the network slices in BS1 so they can be accommodated in the terrestrial wired or wireless links. In this situation, we had only three slices in BS1, with the traffic profiles shown in Fig. 10. The latency requirements are as indicated previously in Table 6.

As stated before, the traffic profile of each slice has 96 samples, from which, we used 67 samples for training, 9 samples for cross validation, and 20 samples for testing. Thus, for the 3 considered slices in this simulation, every training episode has 3 slices $\times$ 67 samples = 201 samples. The cross validation uses 3 slices $\times$ 9 samples = 27 samples, and the testing is performed over 3 slices $\times$ 20 samples = 60 samples.

Since our reward model assigns a reward $+1$ when each slice is allocated, and a reward 0 otherwise, and we confirmed by exhaustive search that all slices can always be allocated, the maximum undiscounted accumulative reward during a training episode is 201, while the maximum undiscounted reward that can be collected during cross validation and testing phases are 27 and 60, respectively.

We trained our DDQN agent for several episodes. The training was stopped when the moving average reward of the episodes, considering a window size of 10 episodes, reached 97.5% of the optimum value, i.e., $0.975 \times 201 = 195.975$.

We repeated the simulations using a neural network with 1, 3 and 5 fully-connected hidden layers, ReLU activations, and 16, 24, 32, 40 and 48 neurons per hidden layer. The number of episodes necessary to achieve the target episode reward (i.e. 97.5% of the optimum) is given by Table 9.

We observe from Table 9 that three of the simulations (highlighted with bold) achieved the highest value (26) during the cross validation, and all these three achieved the optimum result (60) during testing. In these simulations, the cross validation failed in one time step (t = 18:00, slice eMBB), because the wired link achieved exactly the maximum throughput (1 Gbps) and the latency became infinity according to our assumed M/D/1 queuing model [33], violating the latency constraint of the slice.

As shown in Table 9, the simulations with the best cross validation were obtained using critics with 3 or 5 hidden layers, with 40 or 48 neurons per layer. To select the best of these three simulations, several strategies could be followed: 1) choosing the model with least number of parameters to ensure the generalization of the model to new data; 2) choosing the model with the fastest convergence time; or 3) choosing a model with best trade-off between complexity and convergence time. We opted for the third strategy, i.e. among the configurations with the best cross validation results, we select the configurations with the best testing results and from these we choose a configuration with a low number of neurons and training episodes. So, the best results were achieved by the DDQN agent with a critic with 3 hidden layers and 40 neurons per layer, which needs 50 episodes to be trained. This is indeed the simplest model of the three, too, ensuring the generalizability of the model as well.

Fig. 11 illustrates the training episode reward obtained by the DDQN agent when the critic function is composed of 3 fully connected hidden layers with 40 neurons per hidden layer. The figure contrasts the result against 1) the optimum strategy (i.e., reward of 201) obtained by the exhaustive search method; and 2) the random strategy that randomly selects a backhaul link for every slice. We observe that after 50 training episodes, the performance of the DDQN agent approaches the optimum value, while the random approach attains an episode reward much lower than the optimum value.

#### 2) AGENT SELECTS A LINK AMONG TERRESTRIAL AND SATELLITE LINKS, BASED ON THROUGHPUT AND DELAY CONSTRAINTS

In these simulations, we use the action and observation format described previously in Table 3 and Table 4, respectively.

To justify the need for satellite backhaul, we now use four slices (instead of three in the previous simulation), with the traffic profiles depicted in Fig. 8. As shown in that figure, in these simulations, we divided the dataset as previously indicated, with a subset geared for training, cross validation and testing. As we now consider 4 slices, the optimum reward value during training should be 4 slices × 67 samples = 268, the optimum reward during cross validation should be 4 slices × 9 samples = 36, and during testing should be 4 slices × 20 samples = 80. However, using exhaustive search methods, we discovered that, even using satellite, due to the latency constraints, it was impossible to allocate one of the slices in 6 time steps during training, 3 time steps during cross validation, and 5 time steps during testing. Therefore, the best rewards that the DDQN agent could reach were

**TABLE 9.** Results (Throughput and delay; No satellite).

| No. of hidden layers | Neurons per layer | Training episodes | Last episode reward | Average episode reward | Validation reward | Testing reward |
|---|---|---|---|---|---|---|
| 5 | 48 | 41 | 201 | 196.9 | 25 | 60 |
| **5** | **40** | **32** | **196** | **196.0** | **26** | **60** |
| 5 | 32 | 23 | 200 | 196.4 | 23 | 60 |
| 5 | 24 | 124 | 198 | 196.5 | 22 | 59 |
| 5 | 16 | 35 | 197 | 196.3 | 24 | 60 |
| **3** | **48** | **45** | **199** | **196.9** | **26** | **60** |
| **3** | **40** | **50** | **199** | **196.5** | **26** | **60** |
| 3 | 32 | 42 | 198 | 196.0 | 24 | 60 |
| 3 | 24 | 40 | 200 | 196.2 | 20 | 60 |
| 3 | 16 | 29 | 200 | 196.7 | 24 | 60 |
| 1 | 48 | DNC | ---- | ---- | ---- | ---- |
| 1 | 40 | DNC | ---- | ---- | ---- | ---- |
| 1 | 32 | DNC | ---- | ---- | ---- | ---- |
| 1 | 24 | DNC | ---- | ---- | ---- | ---- |
| 1 | 16 | DNC | ---- | ---- | ---- | ---- |

DNC: did not converge to 97.5% of maximum average reward (201)

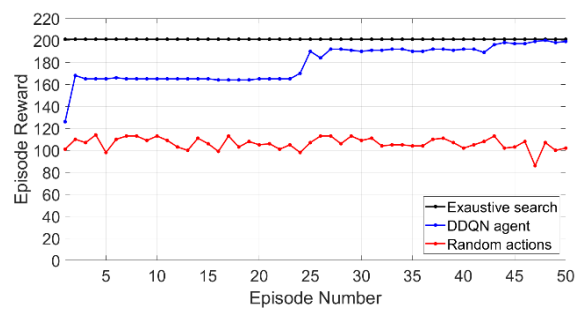

**FIGURE 11.** Evolution of the episode reward during training for the critic with 3 hidden layers and 40 neurons per layer, compared with the exhaustive search and the random selection methods.

268-6 = 262 during training, 36-3 = 33 during cross validation, and 80-5 = 75 during testing.

The simulations ran until a sufficient number of episodes were simulated that allowed to obtain an average episode reward of 97.0% of the best training value (262), considering an averaging window with size equal to 10 episodes.

We conducted several experiments with 1, 3 and 5 fully connected hidden layers, ReLU activations, and 8, 16, 24, 32, 40, 48, 56, 64, 80, 128 and 256 neurons per hidden layer. The number of episodes necessary to achieve the target episode reward (i.e. 97.0% of the best training reward) is given by Table 10.

We can see from Table 10 that several simulations (highlighted with bold) achieved the maximum cross validation reward (33). To select the best of these simulations, we choose once again, from the configurations achieving the best cross validation, the configurations with the best testing results, and from these we choose the configuration with the best trade-off between complexity and convergence time. So, the best results were achieved by the DDQN agent with a critic network with one hidden layer and 80 neurons per layer which needs 21 episodes to be trained, and achieves a good result (70) during testing. From the models in bold, this model is just slightly more complex than the simplest model in the table

TABLE 10. Results (Throughput and delay; With satellite).

| No. of hidden layers | Neurons per layer | Training episodes | Last episode reward | Average episode reward | Validation reward | Testing reward |
|---|---|---|---|---|---|---|
| 5 | 48 | DNC | ---- | ---- | ---- | ---- |
| 5 | 40 | 24 | 257 | 254.4 | 32 | 70 |
| 5 | 32 | 26 | 256 | 254.2 | 33 | 69 |
| 5 | 24 | 24 | 254 | 254.3 | 33 | 60 |
| 5 | 16 | 16 | 257 | 254.4 | 27 | 67 |
| 5 | 8 | 24 | 254 | 254.5 | 33 | 70 |
| 3 | 64 | 23 | 254 | 254.2 | 33 | 69 |
| 3 | 56 | 21 | 256 | 254.4 | 31 | 75 |
| 3 | 48 | 21 | 257 | 254.3 | 33 | 69 |
| 3 | 40 | 22 | 259 | 255.0 | 33 | 70 |
| 3 | 32 | 28 | 258 | 254.8 | 33 | 69 |
| 3 | 24 | 25 | 259 | 254.6 | 33 | 68 |
| 1 | 256 | 26 | 255 | 254.4 | 27 | 70 |
| 1 | 128 | 23 | 254 | 254.6 | 27 | 70 |
| 1 | 80 | 21 | 257 | 254.3 | 33 | 70 |
| 1 | 64 | 42 | 252 | 254.9 | 33 | 70 |
| 1 | 56 | 25 | 256 | 254.3 | 32 | 74 |
| 1 | 48 | DNC | ---- | ----- | ---- | ---- |

DNC: did not converge to 97% of maximum average reward (262)

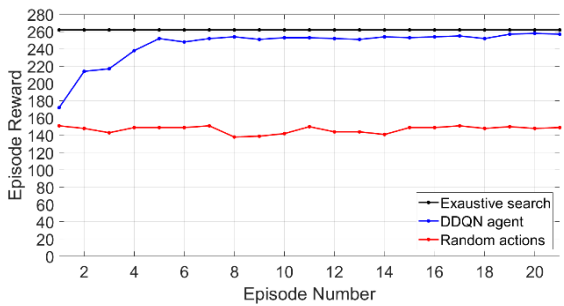

FIGURE 12. Evolution of the episode reward during training for the critic with one hidden layer and 80 neurons, compared with the exhaustive search and the random selection methods.

(i.e., with one hidden layer and 64 neurons per hidden layer), minimizing the risk of overfitting.

Fig. 12 presents the training episode reward obtained by the DDQN agent when its critic function is composed of one fully connected hidden layer with 80 neurons. The figure contrasts the result against 1) the optimum strategy (i.e., reward of 262) obtained by the exhaustive search method, and 2) the random method that selects the backhaul link for each slice randomly. As can be seen, after 21 training episodes, the performance of the DDQN agent approaches the optimum value, while the random strategy attains an episode reward much lower than the optimum value.

Using the same critic (one hidden layer with 80 neurons), we obtained the training results of the agent as shown in Fig. 13, considering the traffic of all slices both in UL and DL. The figure shows the attained backhaul throughput and the failures of the agent due to the violation of QoS requirements of the slices (delay and throughput) over different training episodes.

Analyzing Fig. 13, we see that the DDQN agent learns rapidly in just five episodes, remaining almost stable in the following episodes. This means that the agent learned successfully that:

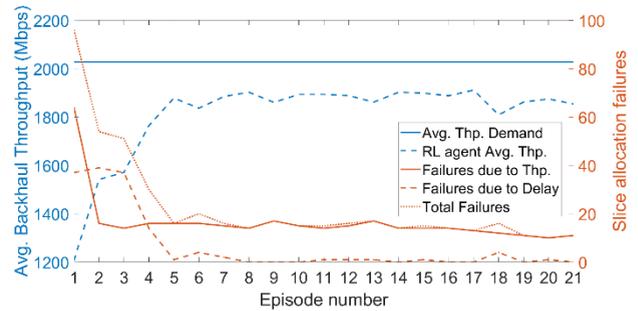

FIGURE 13. Training results achieved by DDQN agent using a critic with one hidden layer and 80 neurons. These results consider the aggregation of the traffic of all four slices in both UL and DL.

- Slice no. 1 (eMBB) could not be allocated to the satellite backhaul because it requires a latency below 100 ms (see Table 6) while the satellite backhaul imposes latencies of 200 ms (see Fig. 5).
- Slice no. 3 (uRLLC) had to be allocated to the wired backhaul because it requires a latency below 1 ms (see Table 6) and the wired backhaul is the only option which imposes lower latencies, i.e. 0.1 ms (see Fig. 5).

However, Fig. 13 also shows that the DDQN agent was unable to accommodate the entire traffic demand in any episode because of failing to allocate some slice in some time steps. This was due to the facts that 1) during six time steps, the network is unable to meet the QoS requirements of one of the slices, and 2) the agent failed to allocate all the slices in at least four of the remaining time steps where the allocation was possible. Moreover, we observe that, except in episodes 2 to 4, the agent failures are mostly caused by the failing in the throughput goals.

### D. DISCUSSION OF RESULTS

Our work studied the problem of selecting a backhaul link to provide additional backhaul capacity in three different scenarios.

The first scenario, i.e., the problem of selecting a wireless backhaul link from a group of terrestrial IAB links, considering only the throughput constraints of the slices, was studied in our previous work [34]. This problem seems to be simple since even a neural network with one hidden layer was able to achieve a quite good result. The best critic configuration, selected among those achieving the best cross validation result and considering the best trade-off between critic complexity and convergence time, was a critic with one hidden layer and 32 neurons, which needs 21 episodes to train (using learning rate $10^{-4}$) [34].

In the second scenario, we verified that the problem of selecting a backhaul link from a group of terrestrial links, satisfying both the throughput and delay constraints imposed by the network slices, was more complex than the previous scenario where we considered only the throughput requirement. As a result, in this problem, as shown in Table 9., the best critic configuration is more complex and requires 3 hidden layers and 40 neurons per hidden layer, which trains in 50 episodes (with a learning rate of $10^{-4}$).

In the third scenario, we added the possibility to borrow backhaul capacity also from a satellite link. We concluded that the problem of selecting a backhaul link, from a group of terrestrial links and a satellite link, that satisfies both the throughput and latency constraints imposed by the network slices can be solved by the DDQN agent using a critic with one hidden layer and 80 neurons that is trained in 21 episodes (using a learning rate of $10^{-3}$).

It is worth noting that given the proposed design of the state and action vectors, our DRL algorithm does not support multi-hop backhaul links. To enable such feature, a multi-agent DRL framework might be incorporated, where each base station is represented by one DRL agent.

## V. CONCLUSION

In this paper, we proposed to build a dynamic wireless backhaul network, comprising IAB and satellite links, that are capable to provide additional backhaul capacity to offload traffic where the wired backhaul capacity becomes momentarily saturated. To achieve this goal, we used machine learning and DRL to construct a DDQN agent that at each time instant and for each slice of the congested base station, selects the additional backhaul link from the pool of available IAB and satellite links. In particular, the studied use cases assumed the use of both wireless IAB and satellite links for backhauling. As for the critic, we conducted experiments with a fully-connected neural network with different sizes to find the critic model with the best trade-off between complexity and convergence time that was able to select the most appropriate backhaul links. From the results, summarized in Tables 9 and 10, we conclude that around 20-50 episodes are sufficient to train a DDQN agent to select the best backhaul links meeting the throughput and latency requirements of all served slices.

For future work, we intend to extend the work to 1) the practical implementation of the proposed scheme in real testbeds; and 2) multi-agent DRL to reduce the communication and computation overheads in large networks.

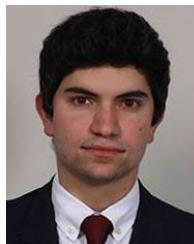

**PABLO FONDO-FERREIRO** received the bachelor's and master's degrees in telecommunication engineering from the University of Vigo, Spain, in 2016 and 2018, respectively, and the Ph.D. degree in information and communication technologies from the University of Vigo, funded by a "la Caixa" Foundation Fellowship, in 2022. In October 2022, he joined the Instituto de Telecomunicações, Aveiro, Portugal, for a two-year postdoctoral research stay. He is currently a Postdoctoral Researcher with the University of Vigo funded by the Xunta de Galicia Fellowship. He has participated in ten competitive research projects and co-authored more than ten papers in peer-reviewed journals and international conference proceedings. His research interests include SDN, mobile networks, MEC, and artificial intelligence. He received the Award for the Best Academic Record.

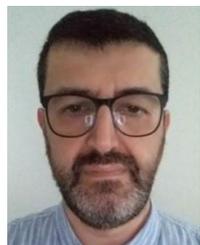

**ANTÓNIO J. MORGADO** received the degree in electronic and telecommunications engineering from the University of Aveiro, Portugal, in 1997. He is currently pursuing the Ph.D. degree with the Faculty of Computing, Engineering and Science, University of South Wales, U.K. Since 1997, he has been with the Instituto de Telecomunicações (IT-Aveiro), Portugal, where he was involved in several Portuguese and European (FP5 and FP6) research projects. He was a Teaching Assistant with the University of Aveiro, from 2003 to 2011, and the Polytechnic Institute of Viseu, Portugal, from 2011 to 2012. He returned to IT-Aveiro, in 2012, to perform research on TV white spaces. From 2014 to 2016, he participated in the European FP7-ADEL project dealing with licensed shared access (LSA). From 2018 to 2021, he was involved in the project H2020-5GENESIS dealing with 5G experimentation testbeds. His current research interests include spectrum regulation, machine learning, 5G/6G standardization, radio resource management, D2D, and mmWave communications.

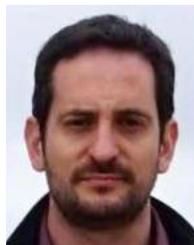

**FELIPE GIL-CASTIÑEIRA** received the M.Sc. and Ph.D. degrees in telecommunication engineering from the University of Vigo, in 2002 and 2007, respectively. From 2014 to September 2016, he was the Head of the iNetS Area, Galician Research and Development Center in Advanced Telecommunications. He is currently an Associate Professor with the Department of Telematics Engineering, University of Vigo. He is also the Co-Founder of a university spin-off, Ancora. He has led several national and international research and development projects. He has published over 60 papers in international journals and conference proceedings. He holds two patents in mobile communications. His research interests include wireless communication and core network technologies, multimedia communications, embedded systems, ubiquitous computing, and the Internet of Things.

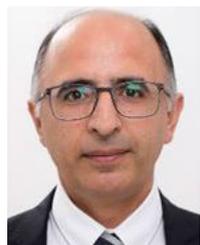

**FIROOZ B. SAGHEZCHI** (Senior Member, IEEE) received the Ph.D. degree in telecommunications from the University of Aveiro, Portugal, in 2016. From 2010 to 2023, he was a Researcher with the Instituto de Telecomunicações, Aveiro, Portugal, where he contributed to several EU-funded research projects, such as H2020-5GENESIS, ECSEL-JU SEMI40, FP7-C2POWER, CATRENE-MobiTrust, and ENIAC-JU E2SG. He is currently a Senior Researcher with the Chair for Distributed Signal Processing (DSP), RWTH Aachen University, Aachen, Germany. His research contributes to the 6GEM project (6G research hub for open, efficient, and secure mobile communications systems), funded by the German Federal Ministry of Education and Research (BMBF). He has authored one editorial book and more than 50 peer-reviewed journal articles and conference papers. His current research interests include mmWave mobile communications, machine learning, MIMO signal processing, and radio resource allocation. He is the Winner of the MDPI Electronics 2024 Best Paper Award. He is an Associate Editor of IEEE ACCESS.

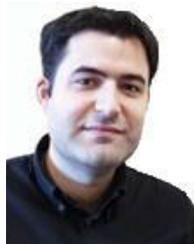

**JONATHAN RODRIGUEZ** (Senior Member, IEEE) received the master's and Ph.D. degrees in electronic and electrical engineering from the University of Surrey, U.K., in 1998 and 2004, respectively. Since 2005, he has been a Senior Researcher with the Instituto de Telecomunicações, Aveiro, Portugal. Since 2017, he has been a Professor of mobile communications with the University of South Wales, U.K. He was the Project Coordinator of the FP7 C2POWER, CELTIC-EUREKA LOOP, and H2020-MCSA-ITN-SECRET. He has carried out consultancy for major manufacturers participating in DVB-T/H and HSUPA standardization. He is the author of more than 600 scientific publications. His research interests include 5G/6G architectures, spectrum management, and radio resource management. He is a fellow of IET (FIET). He has served as the General Chair for several prestigious conferences and workshops. He is a Chartered Engineer (C.Eng.).

∙ ∙ ∙